\documentstyle[draft,psfig]{mn}

\part*{}

\title[SCUBA]{SCUBA: A common-user submillimetre camera operating on the
James Clerk Maxwell Telescope} 

\author[Holland et al.] {W.S. Holland$^1$\thanks{E-mail address:
wsh@jach.hawaii.edu}, E.I. Robson$^1$, W.K. 
Gear$^{2}\thanks{Present address: Mullard Space Science Laboratory,
University College London, Holmbury St. Mary, Dorking, Surrey RH5 6NT,
UK}$, C.R. Cunningham$^2\thanks{Revised address: UK Astronomical
Technology Centre, Royal Observatory, Blackford Hill, Edinburgh EH9 3HJ, 
UK}$, 
\\ \\ \LARGE J.F. Lightfoot$^{2\ddag}$, T. Jenness$^1$, 
R.J. Ivison$^{3}\thanks{Present address:
Department of Physics \& Astronomy, University College London, Gower
Street, London WC1E 6BT, UK}$, J.A. Stevens$^{1\dag}$, P.A.R. Ade$^4$, 
\\ \\ \LARGE  M.J. Griffin$^4$, W.D. Duncan$^{2\ddag}$, J.A. Murphy$^5$
and D.A. Naylor$^6$
 \\ \\
$^1$Joint Astronomy Centre, 660 N. A`oh\={o}k\={u} Place, University
Park, Hilo, Hawaii, USA \\
$^2$Royal Observatory, Blackford Hill, Edinburgh EH9 3HJ, UK \\
$^3$Institute for Astronomy, Department of Physics and
Astronomy, University of Edinburgh, Blackford Hill,
Edinburgh EH9 3HJ, UK \\
$^4$Department of Physics, Queen Mary and Westfield College, Mile End
Road, London E1 4NS, UK \\
$^5$Physics Department, St. Patrick's College, National
University of Ireland, Maynooth, Co. Kildare, Ireland \\
$^6$Department of Physics, University of Lethbridge, Lethbridge,
Alberta T1K 3M4, Canada}

\date{Accepted 1998 September 9; Received 1998} 
\pagerange{000--000} 

\def\laTeX{L\kern-.36em\raise.3ex\hbox{a}\kern-.15em    
T\kern-.1667em\lower.7ex\hbox{E}\kern-.125emX} 

\begin{document} 
\label{firstpage} 
\maketitle 

\begin{abstract}

SCUBA, the Submillimetre Common-User Bolometer Array, built by the Royal
Observatory Edinburgh for the James Clerk Maxwell Telescope, is the
most versatile and powerful of a new generation of submillimetre cameras. 
It combines a sensitive dual-waveband imaging array with a three-band
photometer, and is sky-background limited by the emission from the Mauna
Kea atmosphere at all observing wavelengths from 350\,$\mu$m to 2\,mm. 
The increased sensitivity and array size mean that SCUBA maps close to
10,000 times faster than its single-pixel predecessor (UKT14). SCUBA is a
facility instrument, open to the world community of users, and is provided
with a high level of user support.  We give an overview of the instrument,
describe the observing modes and user interface, performance figures on
the telescope, and present a sample of the exciting new results that have 
revolutionised submillimetre astronomy.

\end{abstract} 


\section{Introduction}

It can be claimed with some justification that the submillimetre waveband
is the last to be opened up for ground-based astronomical research.  Even
from superb sites the atmospheric transparency is generally poor, and even
in the transmission `windows' the high background photon power and
associated noise fluctuations from the atmosphere limit the observing
sensitivity. However, in terms of the three fundamental factors that
govern the impact of a particular waveband -- the collecting area of the
telescope, the sensitivity of the detector, and the availability of
imaging systems -- submillimetre astronomy has finally come of age. 

The 15\,m James Clerk Maxwell Telescope (JCMT) is situated at a high, dry
site--Mauna Kea in Hawaii. With detectors cooled to well below 1\,K,
sky-background noise sensitivity levels are achievable in all the
submillimetre atmospheric windows. Single-pixel instruments such as UKT14
(Duncan et al. 1990), SCUBA's (successful) predecessor on the JCMT,
provided a wealth of photometric and mapping data, though the latter were
limited both in depth and area. Indeed, in terms of extragalactic
astronomy, UKT14 had reached the end of its useful life almost three years
before it was retired from service, with only a handful of high-redshift
active galaxies succumbing to very lengthy integrations in
excellent conditions in the final years (e.g. Dunlop et al. 1994).

\begin{figure*}
\begin{minipage}{175mm}
\psfig{width=155mm,file=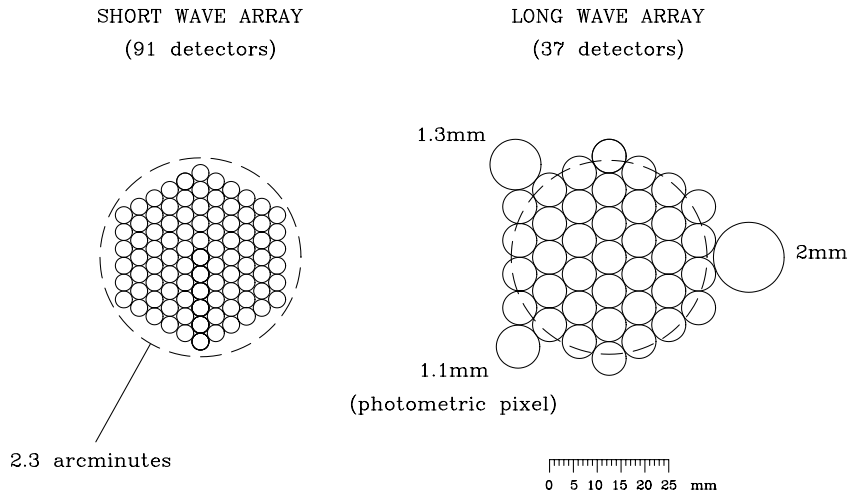,angle=-90,clip=}
\caption{The pixel layout for the SCUBA arrays. The locations of the 
photometric pixels (1.1, 1.35 and 2.0\,mm) are also shown.} 
\end{minipage}
\end{figure*} 

Array detectors were the obvious next step and the extremely ambitious
SCUBA project was approved by the JCMT Board in 1987. SCUBA is not the
first bolometer array to appear on a telescope; over the past several
years there have been the MPIfR 1.3\,mm arrays (Kreysa et al. 1998) on the
Institut de Radio Astronomie Millimetrique (IRAM) 30\,m telescope and the
350\,$\mu$m SHARC array at the Caltech Submillimetre Observatory
(CSO) (Wang et al. 1996). However, SCUBA is unique in combining an
unparalleled sensitivity with an extensive wavelength range and
field-of-view. The result is a mapping capability up to 10,000 times
faster than that of UKT14.
  
SCUBA is a dual camera system containing 91 pixels in the short-wavelength
(SW) array and 37 pixels in the long-wavelength (LW) array. Both arrays
have approximately the same field-of-view on the sky (2.3 arcminutes in
diameter) and can be used simultaneously by means of a dichroic
beamsplitter. The SW array is optimised for operation at 450\,$\mu$m
(but can also be used at 350\,$\mu$m), whilst the LW array is optimised
for 850\,$\mu$m (with observations at 750 and 600\,$\mu$m also
possible). Additional pixels are provided for photometric studies at 1.1,
1.35 and 2.0\,mm. The array pixels are arranged in a close-packed hexagon
as shown in Figure 1, with the photometric pixels positioned around the
outside of the LW array.

SCUBA was delivered from the Royal Observatory Edinburgh to the Joint
Astronomy Centre (JAC) in April 1996 and first light on the
telescope was obtained in July. After extensive commissioning the first
astronomical data for the community was taken in May 1997 using two modes
of operation: photometry and `jiggle-mapping'.  The final major mode of
data-acquisition, `scan-mapping', was released in February 1998. 

SCUBA's forte is the detection of thermal emission from dust with
temperatures ranging from 3\,K to 30\,K. At higher temperatures (except
for high-redshift objects - see later) the dust radiates predominantly in
the far-infrared. SCUBA is also perfectly suited to measuring the
synchrotron emission from extragalactic sources. 

In terms of dust emission, SCUBA has already revealed the population of
galaxies responsible for at least part of the far-infrared background
(Smail, Ivison \& Blain 1997; Barger et al. 1998; Hughes et al. 1998;
Eales et al. 1998), detected a range of high-redshift galaxies (Ivison et
al. 1998a;  Cimatti et al. 1998; Dey et al. 1998), provided new insights
into galaxy evolution (Blain et al. 1998a), imaged the dust extent in
nearby galaxies (Israel, van der Werf \& Tilanus, 1998), discovered
potential proto-planetary systems around a number of `Vega-excess' stars
(Holland et al. 1998a; Greaves et al. 1998), mapped numerous dust
complexes and molecular outflow regions of our Galaxy (e.g. Davis et al.
1998), and discovered new populations of Class 0 protostars and
pre-stellar cores (Visser et al. 1998; Ward-Thompson et al.  1998). The
impact of SCUBA has already been realised and has resulted in an
unprecedented over-subscription factor for the JCMT.

This paper gives a brief description and overview of the SCUBA instrument; 
the detailed technical aspects, as well as the design criteria, will be
found elsewhere (Gear et al. 1998).  We aim to give the user a feel for
the instrument and its operation, the user-interface, software,
performance figures on the telescope, sky noise removal and data
calibration. Finally, we illustrate the potential of SCUBA with a short
selection of new results, and discuss plans for future developments. 

\begin{figure*}
\begin{minipage}{175mm}
\psfig{width=\textwidth,file=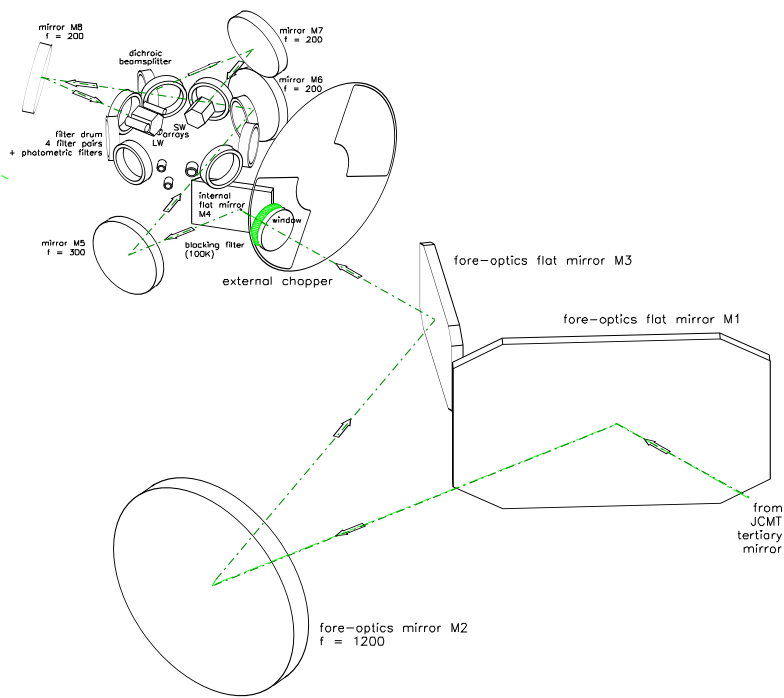,angle=-90,clip=}
\caption{Layout of the SCUBA optics. The dot-dash line shows the
direction of the optical beam through the instrument.} 
\end{minipage}
\end{figure*} 

\section{Optical Design}

SCUBA is mounted on the left-hand Nasmyth platform of the JCMT. To obtain
the maximum field-of-view at this location, the normal $f$/12 focal ratio
has been extended to $f$/16 by a small shift in the secondary mirror
position.  This ensures that the telescope beam passes unvignetted from
the tertiary mirror, through the elevation bearing to the instrument. To
minimise the thermal load on the low-temperature arrays and the size of
the feedhorns, the final focal ratio is down-converted to $f$/4 using a
Gaussian-beam telescope arrangement. This preserves a focal plane that is
frequency invariant, and allows critical components such as the cryostat
window and dichroic beamsplitter to be located at points where the beam
diameter is a minimum. The layout of the optics is shown in Figure 2 and
is fully described in Murphy et al. (1998). 

Just in front of the cryostat entrance window is a sky chopping unit. This
is a dual-bladed chopper that allows the bolometers to view either the sky
(no blade present), the emission from an ambient-temperature blade covered
in Eccosorb, or the reflection of the cold internal optics from the back
of a polished blade. This arrangement provides a standard hot- and
cold-load calibration system similar to existing heterodyne receivers. 

An internal Gaussian-beam telescope relays the input beam to the SW and LW
arrays via a dichroic beamsplitter lying at the final re-imaged aperture
plane; the LW beam being transmitted and the SW beam reflected. The beams
then pass to the arrays through bandpass filters housed in a
rotating filter drum (see section 3). 

A novel feature of the SCUBA optical design is an internal calibration
unit that can remove any short- or long-term drifts in the detector
response (caused, for example, by a change in the sky-background power).
The calibration source is an `inverse bolometer', i.e. a bolometer that
is heated by electrical power and therefore radiates. The signal is
modulated at a frequency of 3\,Hz (i.e. different to the 7.8\,Hz
chop of the secondary) so that both calibrator and chop signals can
be recovered separately. The calibrator is located at an image of the
telescope aperture (mirror M4 in Figure 2) and provides an effective 
point source which uniformly illuminates the arrays through the internal
optics. 

The submillimetre beam is coupled to the detector cavities via conical
feedhorns with single-moded cylindrical waveguides. Each horn illuminates
the telescope with a profile that approximates to a Gaussian, and this
provides efficient coupling to the Airy disk whilst minimising the
response to extraneous background power. The horns are optimised for use
at 450 and 850\,$\mu$m. At the detector, optimum coupling efficiency is
obtained by having a feedhorn with an entrance diameter of about twice the
full-width half maximum (FWHM) of the Airy pattern, i.e. 2$f\lambda$,
where $f$ is the focal ratio of the beam. The array pixels are therefore
spaced at approximately 2$f\lambda$ (ignoring the small wall thickness of
the feedhorn), which results in a significantly undersampled image (the
requirement for Nyquist sampling is $\approx$ $f\lambda$/2). 

To obtain a fully-sampled map at one wavelength requires 4 offsets in
orthogonal directions, and this is achieved by `jiggling' the telescope
secondary mirror over a 16-point pattern to fill in the gaps. This pattern
is repeated as necessary to build up the signal-to-noise (S/N) in a map. 
This mode of mapping is referred to as `jiggle-mapping' (see section 7). 

\section{Wavelength selection}

Ground-based astronomy in the submillimetre waveband is restricted to
observing through particular transmission `windows' in the atmosphere. As
shown in Figure 3 these windows extend from 350\,$\mu$m (860\,GHz) to
850\,$\mu$m (345\,GHz). Throughout this region atmospheric water vapour
is the main absorber of the radiation from astronomical sources. Excellent
observing conditions are therefore also referred to as dry conditions,
meaning the atmospheric water vapour content is low above the site. One of
the best sites currently available is the summit of Mauna Kea, 4.2\,km above 
sea level.

The selection of observing wavelength is made by a bandpass filter
carefully designed to match the transmission window. The filters are
multi-layer, metal-mesh interference filters (Hazell 1991)  located in a
nine-position rotating drum that surrounds the arrays, at a temperature of
around 4\,K. They have excellent transmission (typically over 80\,\%), and
also less than 0.1\,\% out-of-band power leakage. This latter design
characteristic is particularly important as it ensures that there is
minimum contribution to the source signal from extraneous sky emission. 
Composite alkali-halide filters (Hazell 1991) are used to block optical
and near-infrared radiation, thereby minimising the thermal load on the
arrays. These blocking filters are located on the outer radiation shield
of the cryostat (at 90\,K), the filter drum (4\,K) and the 100\,mK array
stage. The filter drum contains several pairs of filters;  for example, it
is possible to observe at 450/850\,$\mu$m simultaneously, and also at
350/750\,$\mu$m. One position of the drum is for the three filters for
the photometric pixels at 1.1, 1.35 and 2.0\,mm. The spectral performance
of the filters was measured by the University of Lethbridge Fourier
Transform Spectrometer (Naylor et al. 1996), and the resultant profiles
are overlaid on the Mauna Kea atmospheric transmission curve in Figure 3.

\section{Bolometers}

In general, bolometric detectors with their large instantaneous bandwidths
have a significant sensitivity advantage over heterodyne receivers. 
However, a drawback of the large bandwidth is an increased susceptibility
to degradation in performance through background power loading (from the
sky and telescope) of the cooled detector (Griffin \& Holland 1988).
Therefore, although bolometers are still the preferred device for
submillimetre continuum astronomy, it is often necessary to restrict the
filter profile to a relative narrow bandwidth ($\lambda$/$\Delta\lambda
\approx$ 5).  This is particularly true for the ultra-low temperature
detectors now in operation. 

\begin{figure}
\psfig{width=85mm,file=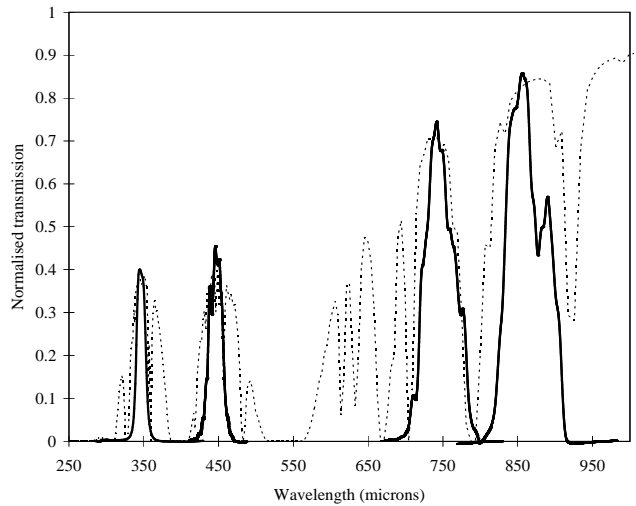,angle=0,clip=}
\caption{The measured SCUBA filter profiles (solid lines), superimposed on
the submillimetre atmospheric transmission curve (dotted) for Mauna Kea
for 1\,mm of precipitable water vapour.} 
\end{figure} 

The SCUBA bolometers are of a composite design. Incoming radiation is
absorbed onto a thin film of bismuth mounted on a saphhire substrate.
Glued to the centre of the substrate is a small chip of neutron
transmutation doped germanium (volume $\approx$ 0.4\,mm$^3$), which acts
as a thermometer and measures tiny changes in temperature due to incident
photons. Electrical connections are made by 10\,$\mu$m diameter brass
wire that govern the overall thermal conductance, and these leads are fed
out of the back of the mount. Each of the 131 bolometers in SCUBA is
designed as a `plug-in' unit that can be close-packed to form an array,
and easily replaced in the event of malfunction (over several years of
extensive testing and operation there has only been one faulty bolometer).
Details of the design and construction of the bolometers are given in
Holland et al. (1996).

When cooled to an operating temperature of 100\,mK and blanked-off to
incoming radiation, the bolometers have a measured electrical noise
equivalent power (NEP) of 6~$\times$~10$^{-17}$\,W\,Hz$^{-1/2}$. Under low
sky-background levels the bolometers achieve background-limited
performance, i.e. optical NEPs of 3.5 $\times$10$^{-16}$ and 1.8
$\times$10$^{-16}$\,W\,Hz$^{-1/2}$ at 450 and 850\,$\mu$m respectively. 
The NEP varies by no more than 20\,\% across each array; this was one 
of the crucial design goals for the instrument. The performance of the
bolometers is fully discussed in Holland et al. (1998b). 
 
\section{Cryogenics}

The SCUBA arrays are cooled to an operating temperature of 100\,mK using
an Oxford Instruments Kelvinox dilution refrigerator. This produces a
cooling power of 20\,$\mu$W at 100\,mK, which is sufficient to cool the
6\,kg of arrays down to this temperature. The refrigerator is built into a
hybrid cryostat consisting of a central 10 litre, liquid helium dewar,
used to cool the filter drum to around 4\,K. A pumped liquid nitrogen
vessel cools the internal optics to 45\,K, minimising the thermal load on
the filter drum and arrays. An outer liquid nitrogen vessel cools a
superinsulated radiation shield and infrared blocking filter to 90\,K,
further minimising the thermal load.

The vacuum system uses a turbomolecular pump that is left running
constantly to limit the migration of contamination onto the arrays. The
vacuum within the instrument is typically less than 10$^{-8}$ mbar. To
minimise resonances and vibrational microphonics within the signal band,
the internal sections of the cryostat are held by fiberglass truss
supports (these also minimise heat conduction). Further details of the
design of the cryogenics can be found in Cunningham et al. (1998). 

\begin{figure*}
\begin{minipage}{175mm}
\psfig{width=\textwidth,file=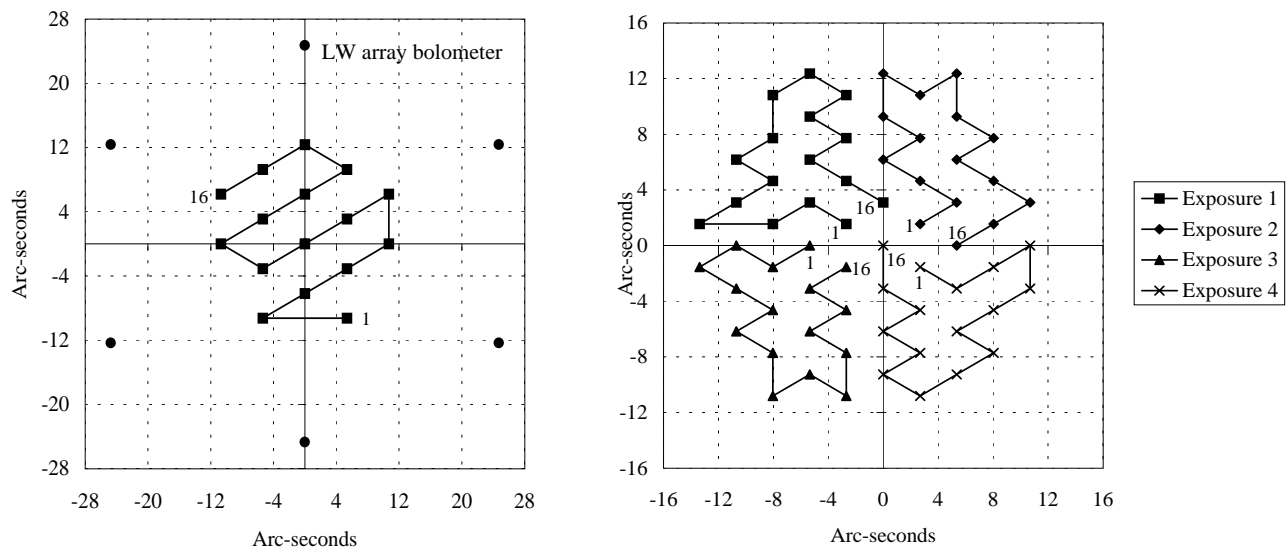,angle=-90,clip=}
\caption{(a) Jiggle pattern to fully sample the LW array alone (b) Jiggle
pattern to fully sample both arrays simultaneously.} 
\end{minipage}
\end{figure*} 

\section{Electronics and signal processing}

The plug-in bolometer units fit directly into a printed circuit board. 
Each bolometer is biased with a 90-M$\Omega$ load resistor (also at
100\,mK) and a common battery voltage supply. The detector signals are
brought to the first stage of signal amplification via superconducting
ribbon cables (Cunningham et al.  1996) that are heat-sunk at various
temperatures inside the cryostat. The JFET head amplifiers are voltage
followers, cooled to 200\,K to minimise gate current noise. The output of
the FETs goes through RF filters in the vacuum seal to external
AC-coupled pre-amplifiers with gains of about 9000. Subsequent to this,
the signals are multiplexed into groups of 16, and pass through an
anti-aliasing filter to reject high-frequency resonances, and finally to
the Data Acquisition System (DAQ).

The DAQ consists of 9 programmable gain amplifiers (to account for
different source flux levels), and the signals are digitised by 12-bit A/D
converters. They then pass to Inmos T800 transputers via a link adapter
and high-speed serial link. The nine transputers perform signal processing
functions such as demodulating the calibrator and chop signals and the
removal of spikes. The image-processing transputer re-samples map data
onto a square grid (usually in RA-Dec) and corrects for sky rotation. Data
are sent via Ethernet to a workstation in the telescope control room for
display. 

\section{Observing modes} 
\subsection{Introduction} 
SCUBA is both a camera and a photometer. There are four basic observing
modes for use with the instrument: photometry, jiggle-mapping,
scan-mapping and polarimetry (the latter needs additional hardware). These
will be briefly discussed in the next section (more extensive descriptions
of the first three are given by Lightfoot et al. 1995; Jenness, Lightfoot
\& Holland 1998). 
 
\subsection{Photometry} 
Photometry is the preferred mode of observation for a completely isolated
point-like source with an accurately known position. Photometry is carried
out with the central pixel of each array simultaneously, or with any of
the long-wavelength photometers independently. The conventional techniques
of two-position chopping and nodding are employed to remove the dominant
sky background. Extensive tests have shown that a better S/N is obtained
by performing a small 3 $\times$ 3 grid (of spacing 2\,arcsecs)  around
the source. This compensates for the slight offset between the arrays
(1.5\,arcsec) and helps cancel atmospheric scintillation and/or slight
pointing or tracking uncertainties. Chopping between pixels on the array
(2-bolometer chopping) so that the source is observed continuously during
the chop cycle is also possible for compact sources. However this requires
a non-azimuthal chop, which is known to be less effective in providing sky
cancellation (Church \& Hills 1990). 

\subsection{Jiggle-mapping}
For sources that are extended but still smaller than the array
field-of-view, jiggle-mapping is the preferred observing mode. This mode
is also useful when searching for point-like sources in a blank field. As
mentioned in section 2, the arrangement of the SCUBA bolometers is such
that the sky is instantaneously under-sampled. For an individual array, 16
offset positions are needed to produce a fully-sampled map (as shown in
Figure 4a for the LW array). For dual-wavelength imaging (eg. at 450 and
850\,$\mu$m) it is necessary to carry out a 64 point jiggle pattern with
3 arcsec sampling. This ensures that the area between the LW array pixels
is covered at the resolution required to fully-sample the SW array (the
spacing of the SW bolometers is about half that of the LW). Since the
telescope is nodded to remove the sky background, a 64 point jiggle
therefore requires 128 sec to complete. It is highly likely that the
sky emission will change over such a timescale and the 64 point jiggle is
therefore split into four sets of 16, with the telescope nodding every 16
seconds (see also Figure 4b).  Jiggle-maps can be coadded to
improve the S/N.

For the telescope user, one remaining issue to consider in setting up
photometry and/or jiggle-map observations is the size and direction of the
chop throw. The chop throw and direction are determined primarily by the
morphology or type of the astronomical source being imaged. Chopping is
normally undertaken in azimuth in order to best cancel the sky emission,
but for some source geometries, specific RA or Galactic co-ordinates might
be required. Chop size is typically 120 arcsec in order to chop off the
array, larger chop throws will usually lead to a degradation in the
quality of an observation through reduced sky-background cancellation (see
section 9). Chop throws as small as 30 arcsec have been used successfully
to search for point sources. 

\subsection{Scan-mapping} 
Scan-mapping is used to map regions that are large compared with the array
field-of-view. This is an extension of the raster-mapping technique used
by single-element photometers (e.g. UKT14), where the telescope is scanned
across a region whilst chopping to produce a differential map of the
source. The SCUBA arrays have to be scanned at one of 6 angles to produce
a fully-sampled map (see Figure 5). The traditional method of
raster-mapping involves scanning and chopping in the same direction
(Emerson, Klein and \& 1979), resulting in a map that has the source
profile convolved with the chop. However, restoring the source data by
deconvolving the chop from the measured map produces problems at spatial
frequencies where the Fourier transform (FT) of the chop (a sine wave) is
low or near-zero. This introduces noise into the restored map. To minimise
these effects, data are currently acquired using a revised method first
described by Emerson (1995) where maps of the same region are taken with
several different chop throws and directions (usually three throws,
chopping in both RA and Dec). This ensures that the zeroes of one chop FT
do not coincide with the zeroes of another. The individual maps are
Fourier transformed and coadded, each weighted according to the
sensitivity of the chop and throw, and then transformed back to give the
final image. This method, which is known locally as the $``$Emerson II''
technique, has been shown to produce substantial improvements in S/N over
the original technique (Jenness et al. 1998c). 

In practice, the telescope is scanned at a rate of 24 arcsec per
second, with a chop frequency of 7.8\,Hz, giving 3 arcsec sampling
along a scan length. A substantial overlap of the array (about half the
diameter) between scans is usually adopted to minimise the effects of
field rotation. As with photometry and jiggle-map, successive scans or
maps over the same region can be coadded together to improve the
S/N. 

\subsection{Polarimetry}
Polarisation measurements with SCUBA require additional hardware in the
form of a photo-lithographic analyser to select one plane of polarisation
along with a rotating half-waveplate. Achromatic waveplates have been
developed to allow simultaneous imaging using both arrays (Murray et al. 
1997). Photometry or fully-sampled maps are taken at a number of waveplate
positions, and the amplitude and phase of the resulting sinusoidal
modulation of the signal is used to deduce the degree of linear
polarisation and the position angle.  Corrections are applied to the data
for instrumental polarisation contributions (due to the instrument and
telescope optics, and also the JCMT membrane) and sky rotation. At time of
writing this observing mode has been commissioned in $``$single-pixel$"$
mode, with the prospects of full-imaging polarimetry by the end of 1998.

\begin{figure}
\psfig{width=85mm,file=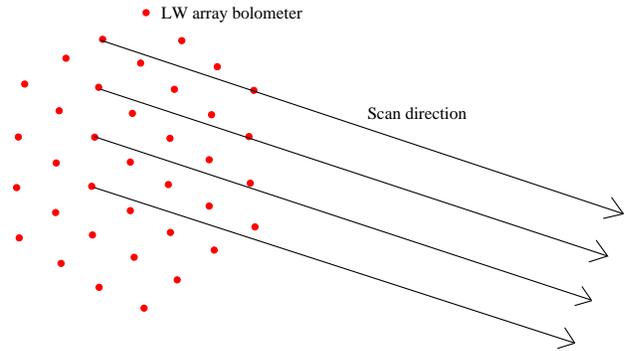,angle=-90,clip=}
\caption{Scan-mapping direction to ensure a fully-sampled map (illustrated
using the LW array).} 
\end{figure} 

\section{The SCUBA queue and flexible scheduling}
SCUBA comes with its own queue software. This enables effective execution
of the observing programme through a series of macros. These are arranged
into an observing queue and their order can be changed at any time (using
observation definition files or ODFs -- see section 10). The queue runs
automatically and when the top observation in the queue is reached, the
telescope drives to the source, the SCUBA parameters are set up and the
observation is performed. This continues without intervention until the
queue is empty. The use of a queue system along with careful pre-planning
goes a long way in maximising the efficiency of the observing process.

As can be seen from Figure 3, even in the transmission windows the
atmosphere is severely attenuating, especially at 450 and 350\,$\mu$m.
Because of the variability of transmission (—and hence detection
performance) with a change of water vapour content, to obtain maximum
efficiency for the facility it is crucial to match the
programme requirements to the weather conditions prevailing at the time. 
Scientific programmes are awarded telescope time by the Time Allocation
Groups (TAGs) with an associated weather flag that specifies the range of
weather conditions in which they can be undertaken.  Those programmes
requiring the very driest conditions are labeled Band 1, whilst
programmes that can be undertaken in relatively wet conditions, such as
bright sources using the 230\,GHz spectral-line receiver, are allocated
to Band 5. 

The programmes are awarded an overall scientific priority and are then
broken down into weather groups at the JCMT. Templates provided by the
Principal Investigator (PI) can then be used to enable the observations to
be undertaken, even without the presence of the PI. This mode of operation
began in February 1998 and a new software suite is currently being
produced to enable visiting observers to identify which programmes should
be undertaken in the prevailing weather conditions according to scientific
priority. This software also provides pipeline data processing, data
tracking and observing programme management tools. The SCUBA observations
represent the first step in the move of the JCMT to a fully flexibly
scheduled facility using weather-based and scientifically prioritised
queues.

\begin{figure}
\psfig{width=85mm,file=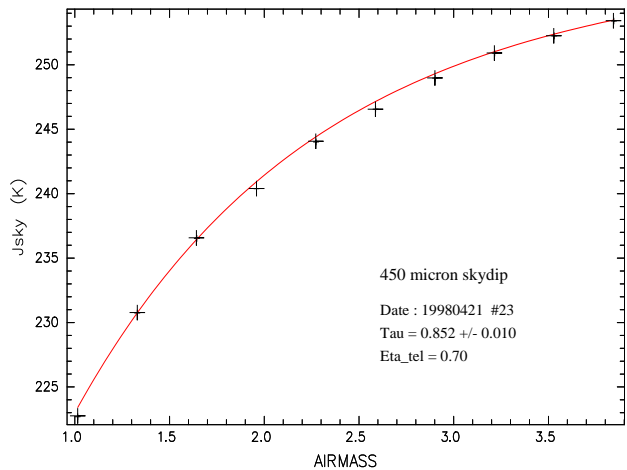,angle=-90,clip=}
\caption{An example of a skydip taken under good observing conditions at 
450\,$\mu{m}$.} 
\end{figure} 

\section{Calibration and sky noise removal}
\subsection{Extinction correction}
The determination of the atmospheric attenuation of a source signal is
critical for the calibration of data in the submillimetre waveband. The
method adopted by SCUBA for measuring the atmospheric transmission is the
technique of skydipping. During a skydip the sky brightness temperature is
measured as a function of elevation (usually between 80 and 15 degrees) 
with absolute temperature reference being provided by a hot-load at
ambient temperature and a cold-load at 45 K. With a knowledge of the
telescope transmission it is then possible to fit the data and derive a
value for the zenith sky opacity (Hazell 1991).  Skydips can be performed
at two wavelengths simultaneously (as for the other observing modes), and
take about 5 minutes to complete. An example of a skydip taken at
450\,$\mu$m is given in Figure 6. It is possible that in the future an
almost instantaneous estimate of the line-of-sight opacity should be
possible from a hot-cold-sky measurement at the elevation of the source,
given a database of sky temperature versus elevation information.

The nearby CSO operates a 225\,GHz radiometer (`the CSO tau-meter') that
performs a skydip in a fixed direction every 10 minutes. The relationships
between the various SCUBA wavebands and the CSO tau-meter are now
well-established (Robson et al. 1998) and this provides additional
assistance in the calibration of data. The CSO tau-meter is used
extensively as an indicator of the conditions before and during an
observing session, and defines the weather-band for a particular night
(and consequently which programmes are to be undertaken).

\subsection{Sky noise and its removal}
In addition to attenuating the source signal, the atmosphere and immediate
surroundings of the telescope contribute thermal radiation that is often
several orders of magnitude greater than the target signal. On many
occasions the primary limit to sensitivity, particularly for deep
integrations at the shorter wavelengths, arises from `sky noise'. 
Sky noise manifests itself in a DC offset and in spatial and temporal
variations in the emissivity of the atmosphere on short timescales.
Chopping and nodding removes the DC offset, but since the chopped beams
travel through slightly different atmospheric paths, the effects of sky
variability are not removed completely. Reducing the chop throw is a
standard and well-tested method of decreasing the effects of sky noise,
and for point-source photometry, chop throws of only 30-40 arcsec are
usually adopted. Increasing the throw to 120 arcsec to ensure chopping off
the array for an extended source, can produce a degradation in S/N of up
to a factor of 2. 

One obvious advantage in using the SCUBA arrays to observe compact sources
(say less than an arcminute or so in diameter) is that there will be a
number of pixels that are viewing the `blank sky' around the source. 
Subtracting the average level of these blank pixels from the source has
been shown to improve the S/N of an observation by up to a factor of 3.
This is illustrated in the mapping observation shown in Figure 7, in which
the output of the central pixel of the LW array at 850\,$\mu$m, and the
average of the outer 18 pixels (`LW sky') are plotted. The noise on each
trace is clearly correlated to a high degree, which results in a
significant improvement in S/N when subtracted (`Centre pixel -- Sky').
Also in this figure is a trace of the SW sky at 450\,$\mu$m (again from
edge pixels)  which demonstrates that sky noise is well correlated between
the arrays. Any number of `sky pixels' can be selected in the data
reduction (usually the more pixels selected the better the S/N)  although
care has to be exercised to ensure there is no `contaminating' source
signal in these pixels. More details of current sky noise removal
algorithms are given in Jenness et al. (1998c). 

\begin{figure}
\psfig{width=77mm,file=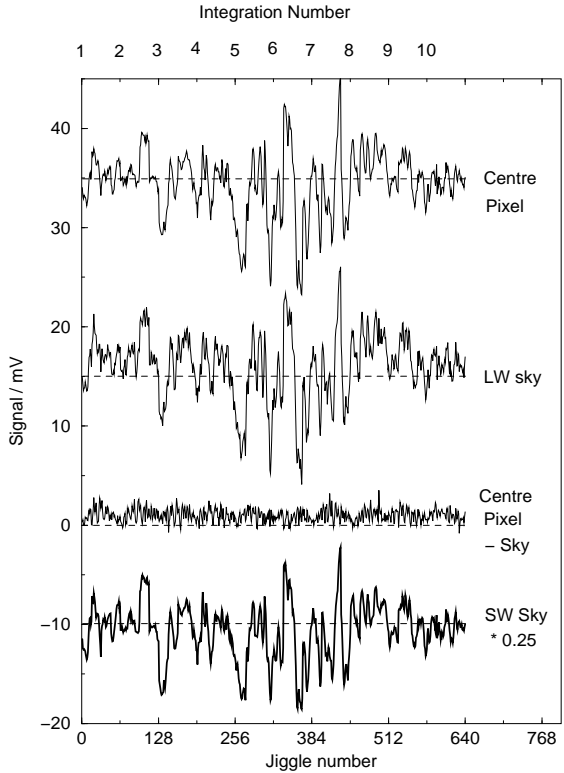,angle=0,clip=}
\caption{Example of the identification and removal of sky-noise.} 
\end{figure} 

Whereas for the photometry and jiggle-mapping modes we have some idea
which bolometers are always looking at sky, for scan-map data the
situation is somewhat different since any bolometer can be on sky {\em or}
source at a given time. This problem is currently overcome by first
removing the source from the data stream before calculating the sky
contribution. This is possible because the source structure should be
constant over time whereas the sky is assumed to vary over the array on
timescales of several seconds. The source signal is removed from the raw 
demodulated data leaving a dataset that should contain only intrinsic sky
emission noise. This sky signature is then removed from the original raw
data, and the image regenerated. An example of the improvement that this
technique generates for scan map data can be found in Jenness et al.
(1998c). One additional advantage of this technique is that it works for
both scan- and jiggle-mapping and removes the need to specify individual
pixels that view sky alone.

\subsection{Calibration}
Primary calibration is taken from Mars (Wright 1976) and Uranus (Orton et
al. 1986). The fluxes of the planets are provided on the JAC
worldwide-web page (Privett et al. 1997) and
the astronomer can use these data along with the beamsizes at the time of
the observation to determine calibration to a high degree of accuracy.
Where planets are not available, a list of secondary calibrators is being
provided by the facility (Sandell et al. 1998). It has been shown to be
important that the calibrators are observed in an mode identical to the
programme sources. Selection of many modes (photometry, jiggle- and
scan-mapping)  during a single shift increases the calibration overheads
significantly. 

\section{Observing tools and the user-interface}
\subsection{Observation definition files}
Prior to programme submission a number of observing planning tools are
available to determine, for example, the necessary integration time to
reach a certain S/N or rms noise level. The use of the Integration Time
Calculator in the application process means that the TAGs have the
correct information to hand, and staff astronomers no longer need go
through the detailed calculations for each application, a notable
saving of staff time. Other tools for determining source position and
optimum time for an observation are also provided.

At the telescope an individual observation, such as a single map or a
single photometric measure of a source, is contained within an observation
definition file (ODF), which is the basic unit of the observing system. 
The ODF specifies everything about the observation, ranging from the
source name (and co-ordinates if it is not in the telescope or observer
catalogue), the wavelength of observation, type of observation
(photometry, mapping), number of jiggles, jiggle spacing, integration
time, number of integrations, etc. Frequently only a small selection of
these parameters require changing by the user, and so it is very easy to
build up a library of ODFs. Special ODFs are also provided for pointing,
focusing and skydips, as well as for calibration. 

The ODFs can be stacked together if needed to make up an observation
macro. This is the usual case where a number of individual measurements
are to be undertaken for a single source. In this case, the macro is
called from the queue and the telescope and SCUBA then perform the ODFs
sequentially until they are complete, or until the user halts or extends
the process to optimise the S/N. An example of a macro for
performing photometry on, say, 3C273, is one that first calls for a
pointing observation of the source at 850\,$\mu$m; the telescope
subsequently centres 3C273 precisely on the central bolometer (usually,
but not necessarily) of the LW array; photometry for a set number of
integrations is then taken at 850/450\,$\mu$m; the filters changed to
perform 750/350\,$\mu$m photometry and, finally, the telescope is offset
to undertake 1.1, 1.35 and 2.0\,mm observations. All this is done
automatically and very efficiently and the data are displayed to the user
in real time. 

\subsection{Real-time display and data reduction}
The use of ODFs and macros makes queue observing very efficient. A 
Telescope System Specialist (TSS) usually monitors the status of the
queue, whilst the observer (or TSS) can insert or re-order ODFs based
on information from the on-line display and the requirements of the
observing programme. The on-line software provides the user with valuable
information concerning the progress of an observation.  For photometry
mode, the output signal from the selected pixel(s) are displayed
graphically, and the S/N of both individual data points and the resulting
coadd are tabulated. For mapping, images are displayed on a re-sampled
RA-Dec plane, and (under stable conditions) the S/N can be seen to build
up in real-time. The on-line display can also be made available to remote
observers via the WORF system (Jenness et al. 1998a). 

Only rudimentary data processing is produced by the on-line display; at
the present time, calibration or sky-noise removal are not applied, and
successive observations of the same source cannot be coadded together. 
However, once an observation is completed, the raw data (stored in
addition to the quick-look data) can be re-processed using the SCUBA User
Reduction Facility (SURF, Jenness and Lightfoot 1998a,b). This allows a
comprehensive analysis of an observation to be undertaken, including
calibration, extinction correction and sky noise removal. In general, it
is possible to keep up with the data reduction, but ideally we would like
the data to be processed by a reduction pipeline as soon as an observation
is complete. For this reason we have recently developed a data reduction
pipeline based on the UKIRT ORAC system (Economou et al. 1998) and SURF. 
This pipeline is currently undergoing testing and should be available at
the telescope in the Autumn of 1998. An additional advantage is that such
a system can be used off-line as well as at the telescope, and so should
make the reduction of SCUBA data almost completely automatic.

\begin{table}
\caption{Measured NEFDs under best and average weather conditions on Mauna
Kea.} 
\begin{center}
\begin{tabular}{@{}ccc}
\hline
Wavelength   & Best NEFD         & Average NEFD          \\
($\mu$m)     & (mJy/$\sqrt{Hz}$) & (mJy/Hz$^{1/2}$)     \\
\hline
350          & 1000              & 1600                  \\
450	     & 450		 & 700			 \\
750  	     & 110		 & 140			 \\
850   	     & 75		 & 90			 \\
1100	     & 90		 & 100			 \\
1350	     & 60		 & 60			 \\
2000  	     & 120		 & 120			 \\
\hline
\end{tabular}
\end{center}
\end{table}

\begin{table}
\caption{5-$\sigma$ detection limits for 1, 10 and 25 hours of
integration time for the best SCUBA NEFDs.} 
\begin{center}
\begin{tabular}{@{}ccccc}
\hline
Wavelength   & NEFD              & 1-hr  & 10-hrs  & 25-hrs  \\
($\mu$m)     & (mJy/Hz$^{1/2}$)  & (mJy) & (mJy)   & (mJy)   \\
\hline
350          & 1000              & 83    & 26      & 17      \\
450	     & 450		 & 38	& 12	  & 7.5     \\
850   	     & 75		 & 6.3	& 2.0	  & 1.3	    \\
1350	     & 60		 & 5.0	& 1.6	  & 1.0     \\
\hline
\end{tabular}
\end{center}
\end{table}

\section{Telescope Performance}
\subsection{Sensitivity}

The overall sensitivity on the telescope is represented by the noise
equivalent flux density (NEFD), the flux density that produces a S/N of
unity in a second of integration. The NEFD, particularly at the shorter
wavelengths, depends very much on the weather, and on many occasions the
fundamental limit to sensitivity is governed by the aforementioned sky
noise (section 9.2).  Table 1 summarises the per-pixel measured NEFD at
all wavelengths after sky noise removal under the $``$best$"$ and
$``$average$"$ conditions at each wavelength. These represent an order of
magnitude improvement over the previous bolometer instrument at the JCMT.
In Table 2 we present 5-$\sigma$ detection limits based on the best NEFDs
for 1, 10 and 25 hours of on-source integration time. 

\begin{figure}
\psfig{width=80mm,file=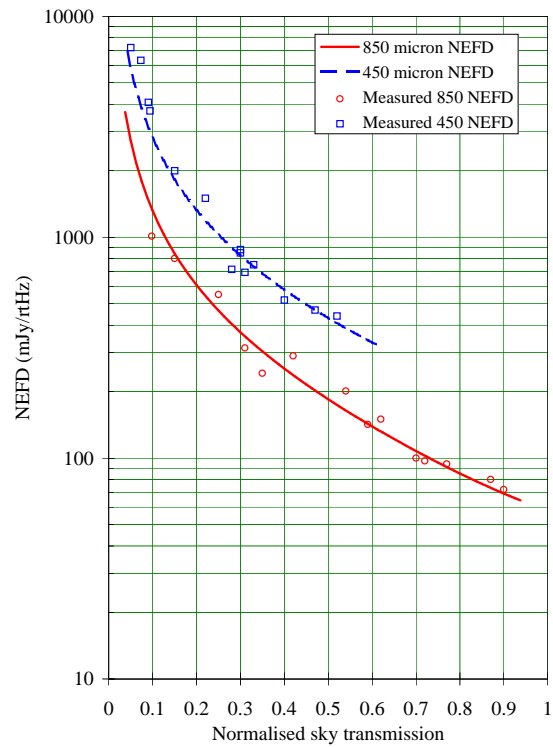,angle=0,clip=}
\caption{Model variation of the NEFD with sky transmission at 450 and
850\,$\mu$m. Measured points are shown by the symbols.} 
\end{figure} 

Since the instrument is predominantly background-limited from the sky, the
NEFD will vary with sky transmission, particularly at the submillimetre
wavelengths. From a knowledge of the components in the optical path
between the detectors and the sky, it is possible to construct a
model of how the NEFD varies as a function of sky transmission (see
Holland et al. 1998b). Figure 8 shows how the calculated NEFDs at 450 and
850~$\mu$m vary with sky transmission.  Also on this plot are some
measured values from long ($\geq$ 15 min) photometry observations,
showing excellent agreement between the model and the measured points.

One of the primary scientific goals for SCUBA is to make deep integrations
of faint sources over a period of many hours, and so it is crucial that
the noise integrates down in a predictable way. Figure 9 shows data
coadded over a period of almost 5 hours, in which the standard error
(solid line) integrates down with time as approximately t$^{-1/2}$ (Ivison
et al. 1998a).

\subsection{Optical performance}

The SCUBA optical system is almost entirely an all-mirror design, and
includes several off-axis mirrors to fold the telescope beam into a
reasonably compact volume. Even though complex mirror shapes minimise
aberrations to a large extent, some field curvature in the focal plane is
inevitable. In addition, the bolometers have a range of optical
responsivities (Holland et al. 1998b). Hence, the positions and relative
responsivities of the bolometers must be accurately known for image
re-sampling to work properly. The arrays are therefore {\em flatfielded}
by scan-mapping the telescope beam over a bright point-like source (e.g. 
Mars or Uranus). The flatfield has been found to remain extremely constant
over time.

\begin{figure} 
\psfig{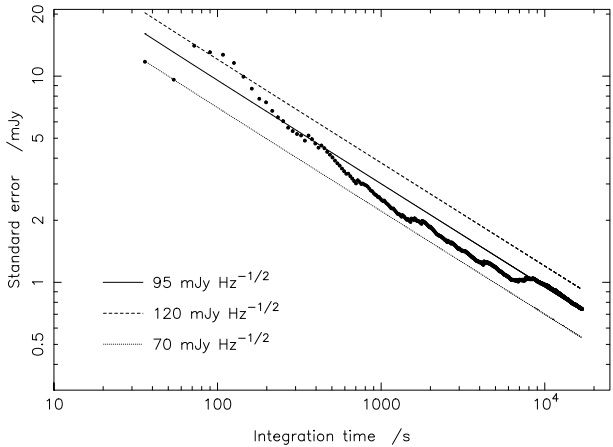}
\caption{Standard error evolution with time for a 5-hour photometry
observation at 850$\mu$m. Three values of system NEFD are plotted and
show that at the time of the observation the NEFD was approximately 95
mJy/Hz$^{1/2}$ (data are from Ivison et al. 1998a).} 
\end{figure}

Diffraction-limited performance is illustrated at 450 and 850\,$\mu$m in
Figure 10. These maps are centred on each array and are a jiggle map of
Uranus, chopping 60 arcsec in azimuth (E-W in the figure). The measured
FWHM beam sizes are 7.8 and 13.8 arcsec at 450 and 850\,$\mu$m
respectively.

\begin{figure}
\psfig{width=75mm,file=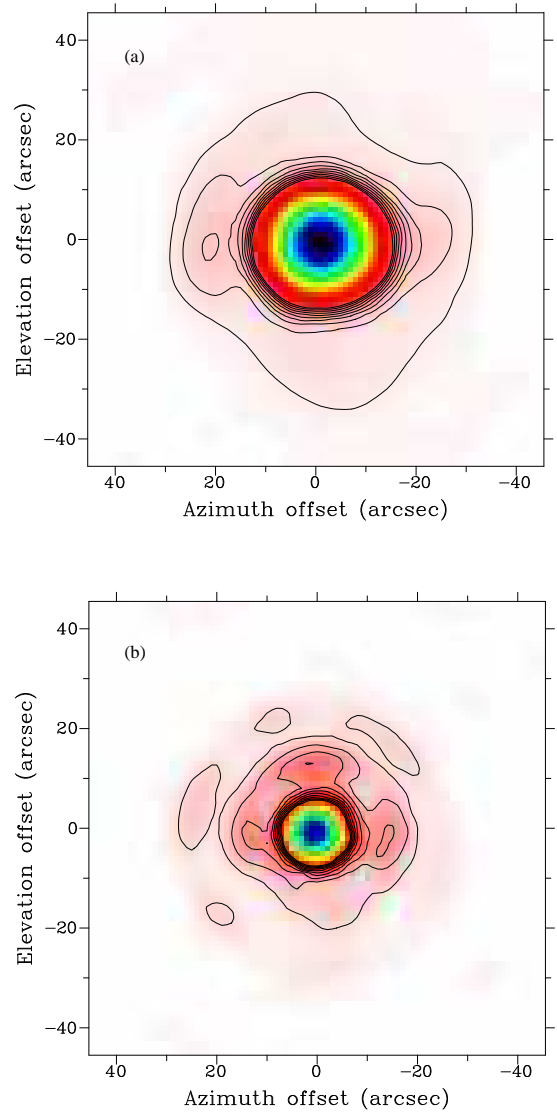,angle=-90,clip=}
\caption{Beam maps of Uranus for the two array central pixels at (a) 
850\,$\mu$m (10 contours starting at 1\,\% of the peak for the base,
1\,\% intervals and (b) 450\,$\mu$m (2\,\% base, 2\,\% intervals).} 
\end{figure} 

\section{Scientific Results}

A number of results from SCUBA have already been published and
here we illustrate just a few examples of SCUBA's potential. Two
techniques stand out that have brought major breakthroughs in
submillimetre science:  {\em deep imaging} and {\em large-scale mapping}.

An example of a bright 450\,$\mu$m SCUBA jiggle-map image is shown in
Figure 11. This is the Eagle Nebula (M16). Some differences are evident
immediately between the HST optical image and the submillimetre map. Note
the increased thermal emission from the tips of the `fingers', totally
invisible on the optical image, possibly revealing the very earliest
phases of star formation -- the `pre-protostellar phase' (White et al. 
1998). The image is a two-field mosaic that was obtained in about 3 hours
of total observing time (which also included an 850\,$\mu$m image). The
peak flux of the brightest finger is 6\,Jy/beam, and the 1-$\sigma$ noise
in the map is approximately 80\,mJy/beam over an area of 6.5 arcmin$^2$.
The equivalent time for the same S/N using SCUBA's predecessor would have
been over 10,000 hours! 

\begin{figure} \psfig{width=80mm,file=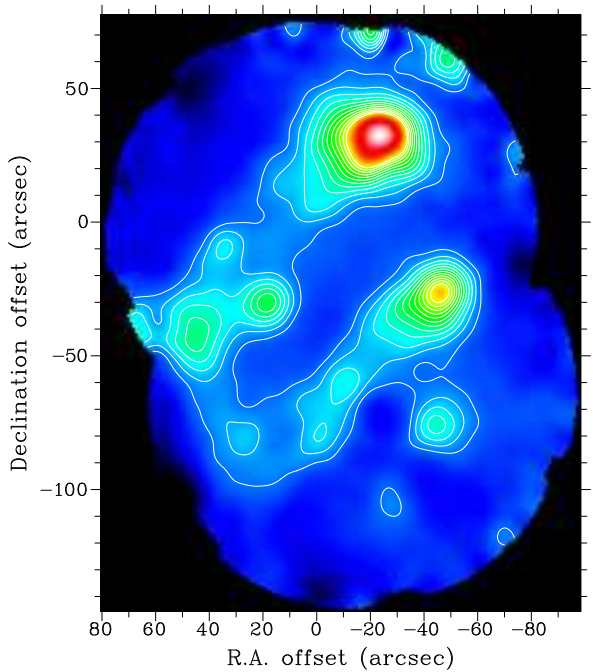,angle=-90,clip=}
\caption{450\,$\mu$m jiggle map mosaic of the Eagle Nebula (M16) (White
et al.  1998). Two of the fingers stand out prominently. Note the excess
emission from the `finger-tips' which is totally invisible on optical and
infrared images.} \end{figure}

One of the long-standing challenges of infrared and submillimetre
astronomy is the understanding of the earliest stages of star formation.
SCUBA is already showing the power of deep imaging to discover new
candidate protostars, as well as obtaining reliable statistics on the
early stages of stellar evolution, including the protostellar Class 0
phase. Recent interest has centred on $``$pre-stellar cores'' which are
significant in that they constrain the initial conditions of protostellar
collapse. Over 40 such cores have already been studied by SCUBA, with more
detections than would be predicted by simple ambipolar diffusion models
(Ward-Thompson et al.  1998). Unbiased surveys of extended dark clouds are
also underway to identify complete samples of protostellar condensations,
allowing the measurement of star formation efficiencies, mass-accretion
rates and evolutionary lifetimes (Visser et al. 1998). 

Observations of circumstellar disks of dust can give clues to the
existence of rocky, Earth-like planets from which they form. SCUBA is
well-suited to measure the low-level thermal emission from dust grains in
such disks. Figure 12 is a 6 hour 850\,$\mu$m jiggle map of the nearby
main-sequence star Fomalhaut. This star was known to have excess
far-infrared emission from IRAS observations, but until the introduction
of SCUBA it had been impossible to image the dust structure in any detail.
The peak flux in the map (28\,mJy/beam) occurs in two distinct peaks,
offset from the star position by about 10 arcsec to the north and south.
The image is consistent with an edge-on torus (doughnut) structure with a
central cavity containing significantly less dust emission. The cavity is
about the diameter of Neptune's orbit, and a possible explanation is that
the region has been cleared of gas and dust by the formation of
planetesimals (Holland et al. 1998a). 

\begin{figure} \psfig{width=85mm,file=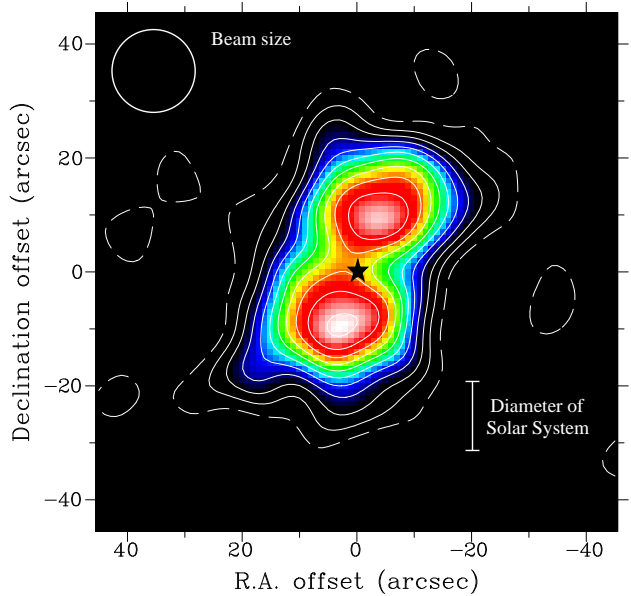,angle=0,clip=}
\caption{850\,$\mu$m jiggle-map of the nearby main-sequence star
Fomalhaut (Holland et al. 1998a). The position of the star is at (0,0) and
is indicated by the $``$star'' symbol. The bar adjacent to the image shows
the apparent diameter of the Solar System (80 AU) if it were located at
the distance of Fomalhaut (7.7 pc).} \end{figure}

The most vigorously star-forming galaxies in the nearby Universe are also
those in which dust obscuration is the most significant. It was long
suspected, therefore, that the early evolution of galaxies would take
place inside shrouds of dust. The first deep SCUBA maps outside the
Galactic plane immediately confirmed this suspicion, revealing a large
population of hitherto unknown, star-forming galaxies. This discovery was
reported by Smail et al. (1997) and subsequent blank-field surveys (Barger
et al. 1998; Eales et al. 1998; Hughes et al. 1998) have confirmed that
the surface density of submillimetre sources is several orders of
magnitude above that expected for a non-evolving galaxy population.
Strongly star-forming galaxies must have been substantially more common in
the early Universe than they are today. The source-confusion limits due to
these dusty galaxies have important consequences for the design and
operation of planned and existing mm/sub-mm telescopes (Blain, Ivison \&
Smail 1998a).

\begin{figure}
\psfig{width=85mm,file=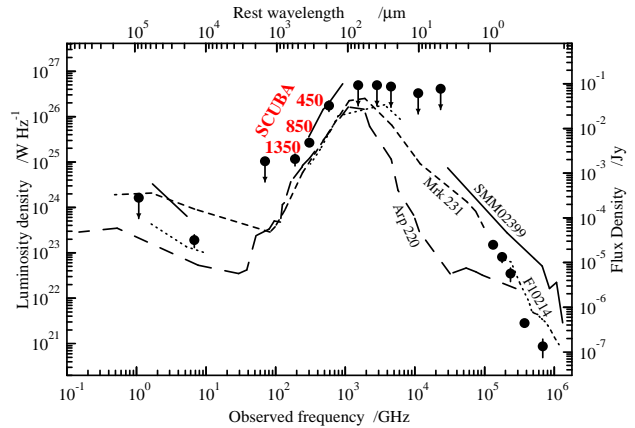,angle=0,clip=}
\caption{The spectral energy distribution of the extremely red object, 
HR10, between the radio and optical (Ivison et al. 1997; Dey et al.\ 1998), 
highlighting the SCUBA detections at 450, 850 and 1350\,$\mu{m}$. For 
comparison, we have also plotted the SEDs of the ultraluminous {\em IRAS} 
sources F\,10214+4724 {\em (dots)} (Rowan-Robinson et al. 1993), Mrk~231 {\em 
(dashes)} (Barvainis et al. 1995), Arp~220 {\em (long dashes)} (D.H. Hughes, 
priv. comm), and the radio galaxy 8C\,1435+635 {\em (solid)} (Ivison et al. 
1998a). The lines are broken in regions where only upper limits are 
available. For F\,10214+4724 they are corrected for lensing by a factor of 30 
(e.g. Broadhurst \& Leh\'ar 1996).}
\end{figure}

\begin{figure}
\psfig{width=87mm,file=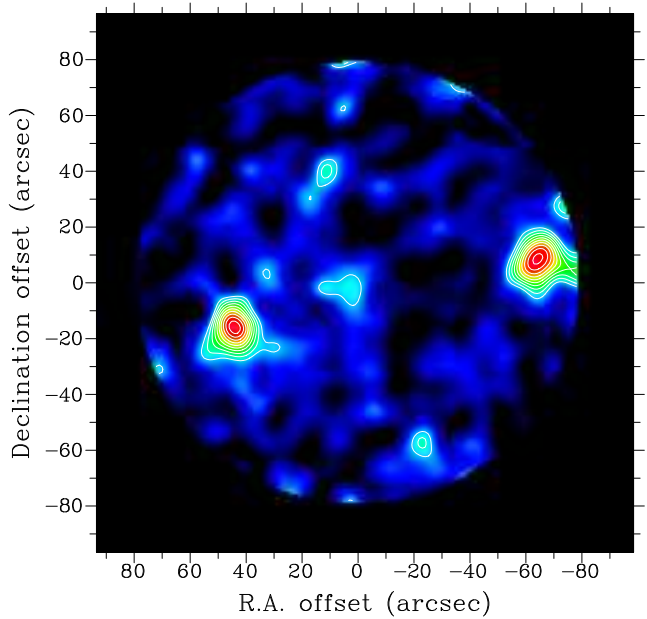,angle=-90,clip=}
\caption{Deep 850\,$\mu$m map of a field in the massive galaxy cluster,
Abell\,1835, at $z=0.25$, showing discrete submillimetre emission from
extremely luminous, dusty, star-forming galaxies behind the cluster
(Smail et al.\ 1998b), and from the central cluster-dominant galaxy
(Edge et al.\ 1998). The bright source to the east is a system of
interacting galaxies at z $\simeq$ 2.6 (Ivison et al. 1998c).} 
\end{figure} 

These first deep submillimetre surveys imply that a large population of
dusty galaxies is missing from optical inventories of star-formation
activity (Blain et al. 1998b). Further support for this was obtained with
the submillimetre detection of an extremely red galaxy, HR\,10, at
$z=1.4$. This is a relatively common class of galaxy previously thought to
consist of very old, quiescent ellipticals, but which SCUBA has revealed
to be young, star-forming systems similar to local ultraluminous {\em
IRAS} galaxies (ULIRGs)  (Cimatti et al. 1998; Dey et al. 1998).  The
distant, submillimetre-selected galaxies discovered by Smail et al. (1997) 
also resemble ULIRGs, at least in the rest-frame ultraviolet/optical, with
a similar proportion of mergers (Sanders \& Mirabel 1996; Smail et al. 
1998a). The spectral energy distribution of HR\,10 is shown in Figure 13,
and an example of a deep 850\,$\mu$m image revealing dusty, extremely
luminous galaxies behind a massive galaxy cluster is shown in Figure
14.

The study of galaxies with active galactic nuclei (AGN) has also been
revolutionised by SCUBA. Use of the jiggle-map mode led to the discovery
of SMM\,02399$-$0136, a hyperluminous galaxy at $z=2.8$ hosting an AGN
(Ivison et al.\ 1998b; Frayer et al. 1998). Such galaxies cannot be easily
detected in conventional AGN/QSO surveys, so the presence of
SMM\,02399$-$0136 in the very first submillimetre image of the distant
Universe may suggest that estimates of the prevalence of AGN may require
substantial revision.

The unprecedented sensitivity of SCUBA's photometry mode has allowed the
study of radio- and optically-selected AGN to move from the pioneering
world of bare detections to the reliable extraction of physical
parameters. For the high-redshift radio galaxy, 8C\,1435+635, Ivison et
al.\ (1998a) presented 450 and 850\,$\mu$m detections of sufficient
quality to infer that the formative starbursts of such massive ellipticals
may still be in progress at $z\simeq 4$. Observations of a complete sample
of radio galaxies spanning a range of redshifts and radio luminosities
will be presented by Dunlop et al.\ (1998). 

Finally, as an illustration of the power of SCUBA to map large areas,
Figure 15 shows a 850\,$\mu$m scan map of the central region of the
Orion Molecular Cloud (OMC). This 8 $\times$ 10 arcmin commissioning map
was obtained in just 50 minutes of integration time, and has a noise level
of about 60\,mJy/beam.  The famous Orion ``Bright-Bar'' is clearly seen in
the image, as well as the little-studied embedded protostellar ridge in
the north. The region is also believed to be a site of progressive star
formation (from the south to the north), and so offers an opportunity to
compare dust core properties over a range of evolutionary stages. This map
is taken from a paper by Greaves and Holland (1998), and would have been
impossible for UKT14 at the JCMT in a reasonable amount of observing time. 
A much larger region of the OMC has since been imaged using the
$``$Emerson II'' technique and is described by Johnstone et al. (1998). 

\begin{figure} 
\psfig{width=80mm,file=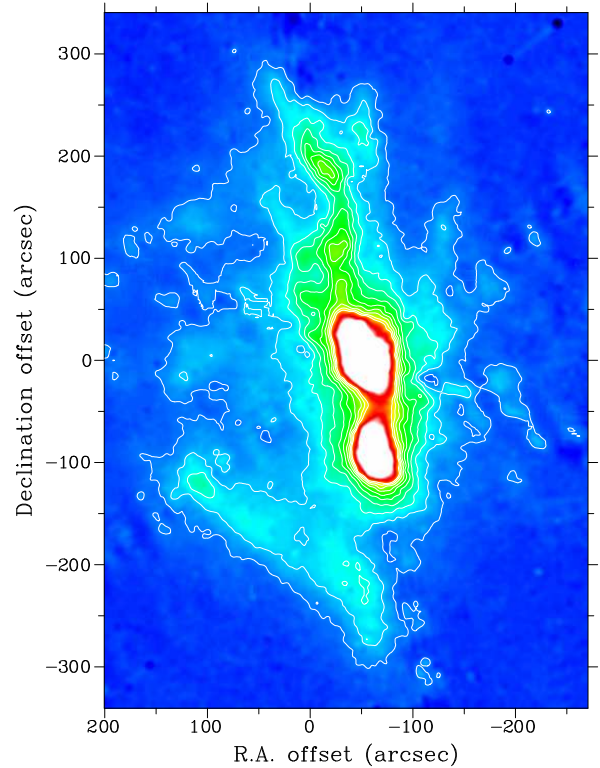,angle=-90,clip=}
\caption{Scan-map of the central region of the Orion Molecular Cloud
(Greaves and Holland 1998). The central cores of OMC1 and OMC1S have been
`overexposed' to highlight the low-level structure of the $``$bright-bar''
and northern ridge.} 
\end{figure}

\section{Current and future developments}

SCUBA is a continually evolving instrument. There are a number of
identified improvements both in terms of observing strategies and
sensitivity gains. These are briefly described in this section. 

\subsection{Observing modes}

A new mode of `fast data sampling' is currently under investigation. Here,
the chopping component and the image sampling are merged together into one
step pattern (Le Poole \& van Someron Greve 1998) which is completed as
rapidly as possible (twice per second) in order to compensate for any sky
variabilities. This is repeated many times to build up the source
integration time. There are two immediate benefits. The removal of the
requirement for a bolometer to look at blank sky for 50\,\% of the time
(during the `off' cycle of the nod), which, for the same reasons as two
bolometer chopping on the array, can in principle bring an efficiency
improvement of a factor of two. Also, because the pattern is very much
faster than a conventional 16\,sec jiggle, there is potential for
real-time reduction of sky noise in the resulting data. This mode is
undergoing telescope testing in mid-1998.

It is generally agreed that the best method for sky-noise suppression is
to use three-position chopping (Papoular 1983). This works best for
compact sources where the two `off' beams are themselves sampled by other
bolometers on the array. In principle, this technique should give a factor
of $\sqrt{2}$ improvement in noise. Tests are ongoing at the telescope and
it is likely that this mode will be released early in 1999.

\subsection{Sensitivity improvements}

Although the sensitivity per SCUBA pixel is approximately an order of
magnitude better than the previous instrument, it is expected that there
are still substantial gains to be made. One obvious way to improve the SW
array performance in particular, is by improving the large scale surface
accuracy of the telescope primary mirror. A programme is underway to
accomplish this, and by Autumn 1999 it is expected that the 450\,$\mu$m
NEFD per pixel will improve by a factor of approximately 2 under good
observing conditions.

There are significant discrepancies between the estimated and measured
optical efficiencies at all SCUBA submillimetre wavelengths. The cause of
this loss is thought to be due to a interaction between the array
feedhorns and the filters (which are in close proximity). It is possible
that a large array of feedhorns so close to the filters modifies the
effective terminating impedance of the filters, resulting in a change in
their reflectance coefficients. The net result is that the horn beams
leaving the filter drum enclosure are considerably broader than when
measured individually.  If the beams are sufficiently wide that they
over-illuminate the internal mirrors then not only will the signal
coupling to the telescope be reduced, but additional photon noise will be
introduced from the 45\,K optics box. If the horn beams themselves cannot
be corrected, one possible way of counteracting the rise in photon noise
would be to terminate the spillover on a colder surface. An additional
radiation shield at a temperature of $<$\,10 K would be complex and
difficult to manufacture but is nevertheless being considered. An
alternative solution would be to use a helium liquifier to cool the optics
box to around 4\,K.  Full cost-benefit analyses of these options have yet
to be undertaken; if proven viable it will not be in place before early
2000. 

FTS filter measurements of the photometric pixels revealed a deep fringe 
roughly in the centre of the observed profile, and the depth of the fringe
increased with wavelength. All the bolometers are the same and the
cavities do not scale with wavelength. Consequently, at the longer
wavelengths, where fewer modes propagate, the integrating properties of
the cavity are diminished. A simple `tuning' of the cavity for the
1350\,$\mu$m pixel (moving the waveguide closer to the substrate) gave
an immediate factor of two improvement in NEFD. Such gains should also be
expected at 1.1 and 2.0\,mm, although it is unlikely that there will be
any significant gains at the array wavelengths.

\subsection{Noise performance}

There are some 10\,\% of pixels that do not meet the noise specification
(Holland et al. 1998b). The noise has a characteristic 1/frequency
signature, and is thought to be due to poor contacts in the ribbon cables.
Fortunately, there has not been a deterioration in noise performance (e.g.
due to oxidisation or moisture when the instrument is opened). A new
ribbon cable technology, being developed by the UK ATC, Edinburgh,
should eliminate this problem. New cables are expected in early 1999. 

One of the early problems encountered with SCUBA on the telescope was an
unexpected sensitivity to vibration from the secondary mirror unit (SMU).
This affected about 10\,\% of pixels, in addition to those suffering from
1/f noise. The demodulation of the chop and calibrator signals was
severely affected by the dominant 5th harmonic of the chop at around
40\,Hz. Increasing the risetime of the chopper waveform by about 2\,msec
and providing a smoother transition at the end of the cycle improved the
situation dramatically without a noticeable loss in on-source efficiency
(and consequently NEFD). Although several pixels remain sensitive to SMU
vibration, particularly for large chop throws, this is no longer a major
concern. However, the $``$fast-data sampling'' mode is sensitive to a
wider range of frequencies (0.5 - 30\,Hz) than the conventional observing
modes. Features caused by telescope resonances (induced by drive motors
and compressors) around 25 - 30\,Hz are evident in the bolometer frequency
spectra. This is currently the limiting factor to achieving the efficiency
improvements described in section 13.1. A major mid-life electronics
upgrade to reduce the sensitivity to vibration by using differential
amplifiers is currently being considered.

\subsection{Array upgrades}

Bolometer array technology has advanced quite significantly since the
design of SCUBA. It is possible that SCUBA could be upgraded to take
advantage of the planar arrays which are currently being developed, e.g.
for the SPIRE bolometer instrument aboard the Far-Infrared and
Submillimetre Space Telescope (Griffin et al. 1998). Filled absorber
arrays without feedhorns, and containing perhaps many hundreds of
detectors, have an added potential of a gain of about three in mapping
speed compared to the 2$f\lambda$-spaced SCUBA bolometers. Once the number
of pixels increases much beyond about 200, more novel signal readout
schemes are also needed;  this technology is currently being developed. 

\section{Conclusions}

SCUBA has proven to be an extremely powerful and versatile camera
for submillimetre astronomy. Background photon noise limited performance
on the telescope has been achieved by cooling the detectors to 100\,mK and
careful design of the focal plane optics. With an order of magnitude
improvement in per-pixel sensitivity over the previous (single-pixel)
instrument, and over 100 detectors in two arrays, SCUBA can acquire data
approaching 10,000 times faster than was possible previously. It is clear
that with this huge increase in performance, SCUBA will make big advances
in submillimetre astronomy for many years to come.

\vskip 2mm

\section*{ACKNOWLEDGMENTS}

The JCMT is operated by the Joint Astronomy Centre, on behalf of the U.K. 
Particle Physics and Astronomy Research Council, the Netherlands
Organisation for Pure Research, and the National Research Council of
Canada. RJI is supported by a PPARC Advanced Fellowship. We take this
opportunity to thank all those people involved in the design,
construction, testing and commissioning of SCUBA. The image of the galaxy
cluster, A1835, is reproduced with the kind permission of Ian Smail and
Andrew Blain. Further details and updated information can be found on the
SCUBA world-wide webpage at URL: http://www.jach.hawaii.edu/JCMT/scuba.

\bsp 
\label{lastpage} 
\end{document}